# Dominance of backward stimulated Raman scattering in gas-filled hollow-core photonic crystal fibers

**MANOJ K. MRIDHA\*, DAVID NOVOA, AND PHILIP ST.J. RUSSELL**

*Max Planck Institute for the Science of Light, Staudtstrasse 2, 91058 Erlangen, Germany*
*\*Corresponding author: manoj.mridha@mpl.mpg.de*

**Backward stimulated Raman scattering in gases provides a promising route to compression and amplification of a Stokes seed-pulse by counter-propagating against a pump-pulse, as has been already demonstrated in various platforms, mainly in free-space. However, the dynamics governing this process when seeded by noise has not yet been investigated in a fully controllable collinear environment. Here we report the first unambiguous observation of efficient noise-seeded backward stimulated Raman scattering in a hydrogen-filled hollow-core photonic crystal fiber. At high gas pressures, when the backward Raman gain is comparable with, but lower than, the forward gain, we report quantum conversion efficiencies exceeding 40% to the backward Stokes at 683 nm from a narrowband 532-nm-pump. The efficiency increases to 65% when the backward process is seeded by a small amount of back-reflected forward-generated Stokes light. At high pump powers the backward Stokes signal, emitted in a clean fundamental mode and spectrally pure, is unexpectedly always stronger than its forward-propagating counterpart. We attribute this striking observation to the unique temporal dynamics of the interacting fields, which cause the Raman coherence (which takes the form of a moving fine-period Bragg grating) to grow in strength towards the input end of the fiber. A good understanding of this process, together with the rapid development of novel anti-resonant-guiding hollow-core fibers, may lead to improved designs of efficient gas-based Raman lasers and amplifiers operating at wavelengths from the ultraviolet to the mid-infrared.**

## 1. Introduction

Backward stimulated Raman scattering (BSRS) has been investigated as a route to generate short pulses that, through compression, reach peak powers exceeding those of the parent pump pulses [1-3]. Compared to solid materials, Raman-active gases are suitable for these studies, providing large Raman frequency shifts, narrowband emission lines, high stimulated Brillouin scattering thresholds and high optical damage thresholds [4]. The most common experiments on BSRS involve free-space arrangements in which the seed pulse is first generated by forward stimulated Raman scattering (FSRS) and then launched into a second gas cell in a counter-propagating geometry. The primary goal of most previous work has been to efficiently generate strong backward Stokes signals (frequency downshifted by the Raman transition frequency) or short pulses with high peak power. In free space geometries, analysis of the interplay between FSRS and BSRS is rather complex as it strongly depends on parameters such as pump pulse duration, focusing conditions, interaction length and gas pressure [4-7]. Moreover, the threshold for these processes is usually very high, typically requiring pulse energies in the multi-mJ range. The introduction of wide-bore capillaries reduced to some extent the pump-pulse energies required, although at the expense of high attenuation of the generated signals [8,9].

The advent of hollow-core photonic crystal fiber (HC-PCF), which offers low confinement loss, broad transmission windows, tight confinement of light in cores of few tens of microns in diameter, and pressure-tunable dispersion, has made it possible to reduce the FSRS threshold by orders of magnitude [10,11]. HC-PCFs have been used in, for example, efficient Stokes and anti-Stokes emission, generation of Raman frequency combs [12] and molecular modulation of arbitrary optical signals [13]. As in free space and wide-bore capillaries, pulse compression and amplification of counter-propagating Stokes signals can also be observed in gas-filled HC-PCFs [14]. In our experiments we used hydrogen, which has the highest material Raman gain and frequency shift of any gas, and is transparent down into the ultraviolet.

In this article we report for the first time clear noise-seeded backward SRS in $H_2$-filled kagomé-style HC-PCF, pumping at 532 nm. Using narrowband pulses 3.2 ns in duration (physical length 96 cm) and a 114 cm length of fiber with core diameter 22 μm filled with 38 bar of hydrogen, the noise-seeded backward Stokes signal (at 683 nm) was always stronger than the forward, reaching efficiencies exceeding 40%. Even higher backward conversion efficiency (65%) was obtained when a 14 cm length of fiber was used—in this case the pump pulse is substantially longer than the fiber, so that the backward Stokes process could be seeded by a small fraction of the forward Stokes signal, back-reflected at the window of the gas cell.

To understand the in-fiber temporal dynamics of backward Raman scattering we recorded the temporal profiles of all the interacting sidebands along with their energy and compare the results with numerical simulations based on a bi-directional set of Maxwell-Bloch equations [14, 15, 16]. This allowed us to relate the dynamics of the BSRS process to the evolution of forward and backward Raman coherence waves (i.e., synchronous molecular oscillations driven by the beat-note created by the pump and Stokes signals), which turns out unexpectedly to favor the onset and amplification of the backward signal, even though the actual gain coefficient for FSRS is higher than that for BSRS.

The paper is organized as follows: in section 2 we will briefly revise general aspects of the Raman gain in hydrogen gas. In section 3 we present the experimental set-up employed to demonstrate self- and noise-seeded BSRS in HC-PCF, followed by the discussion of the experimental and numerical results (section 4). In section 5 we discuss the key role of the interplay between forward and backward coherence waves in the dominance of BSRS over FSRS at high gas pressures.

Finally, in section 6 we summarize the conclusions of the present study and give an outlook of the potential implications of the results across different fields.

## 2. Raman gain in molecular hydrogen

The steady-state material Raman gain of the fundamental vibrational mode of $H_2$ (with frequency $\Omega_R \sim 125$ THz) is given by [17]:

$$g_P(\nu_P, p) = \frac{9.37 \times 10^{12}(57.2 p/\Delta\nu)(\nu_P - \Omega_R)}{c(7.19 \times 10^{13} - \nu_P^2/c^2)^2} \quad \text{m/W} \quad (1)$$

where $\nu_P$ [Hz] is the pump frequency, $p$ [bar] the gas pressure, $\Delta\nu = (280/p) + 56.98 p$ is the Raman linewidth [MHz] and $c$ [m/s] is the speed of light in vacuum. In general the steady-state BSRS gain is not equal to the FSRS gain (which depends mainly on $\Delta\nu$) and is strongly dependent on the pump linewidth and Doppler-broadened linewidth [2]. In essence, the BSRS/FSRS gain ratio can be approximated as $R \sim \Delta\nu_f / \Delta\nu_b$, where $\Delta\nu_f$ and $\Delta\nu_b$ are the net linewidths for the forward and backward Raman scattering. At high pressures both $\Delta\nu_f$ and $\Delta\nu_b$ increase under the influence of collisional broadening, and begin to converge ($R \to 1$). Throughout this paper we therefore use $R = 0.7$ in the numerical simulations, based on forward and backward linewidth measurements reported for the fundamental vibrational transition of $H_2$ [18]. Under these conditions in free-space arrangements, the backward Stokes signal has already been shown to compete and even overtake the forward Stokes signal [4, 5, 6, 19, 20]. The large numbers of degrees of freedom and relative complexity in free-space systems has meant, however, that a clear explanation of this phenomenon has not so far been possible.

## 3. Experimental set-up

The experimental set-up is sketched in Fig. 1(a, b). A frequency-doubled Nd:YAG laser operating at 3 kHz repetition rate delivered linearly-polarized 532 nm pump pulses of duration ~3.2 ns FWHM (corresponding a pulse length of ~96 cm in the fiber). Two kagomé-style HC-PCFs with different core diameters and fiber lengths were employed (see Fig. 1(c) for scanning electron micrographs (SEM) of the fiber structures). In the first set of experiments we used a 14-cm-long fiber with ~47 μm core diameter (Fiber 1) and in the second set a 114-cm-long fiber with ~22 μm core diameter (Fiber 2). The first fiber was placed inside a 16-cm-long monolithic gas-cell, while the second fiber was placed in an arrangement of two 8-cm-long gas-cells connected by a ~1 m long tube. Both fibers were filled with 38 bar of hydrogen—the highest pressure attainable with the gas system.

The pump pulses were launched into the fibers using lenses with a maximum efficiency of ~70%. A dichroic mirror (DM1) transmitting the first vibrational Stokes line at 683 nm and fully reflecting the 532 nm pump line was placed in front of the in-coupling lens. The beams emerging at both ends of the fiber were collimated and diverted to different detectors. The energies of the pump and generated forward and backward Raman pulses were measured with a calibrated power meter using customized band-pass filters. An optical spectrum analyzer was used to measure the spectral content of the output signals. The far-field profile of the backward Stokes signal was also recorded with a CCD camera.

Alongside the energy-spectral measurements we performed thorough time-resolved analysis of the generated signals to understand the competing dynamics of FSRS and BSRS. The temporal profiles of the pump, forward and backward Stokes pulses were recorded with fast photodiodes triggered by the pump pulse. The position of these photodiodes was carefully chosen to ensure their synchronism (see Fig. 1(b) for schematics)—a basic requirement for subsequently comparing the temporal evolution of the signals in an unambiguous manner.

In order to check the effects of self-seeding, we specifically designed two different window holders for the gas-cells (see schematics in Fig. 1(a)), mainly for the output end: a flat holder for normal incidence and an angled holder for 40° incidence. On these holders, 3 mm thick $MgF_2$ windows were mounted in such a way that the distance between the fiber output end and the window was preserved.

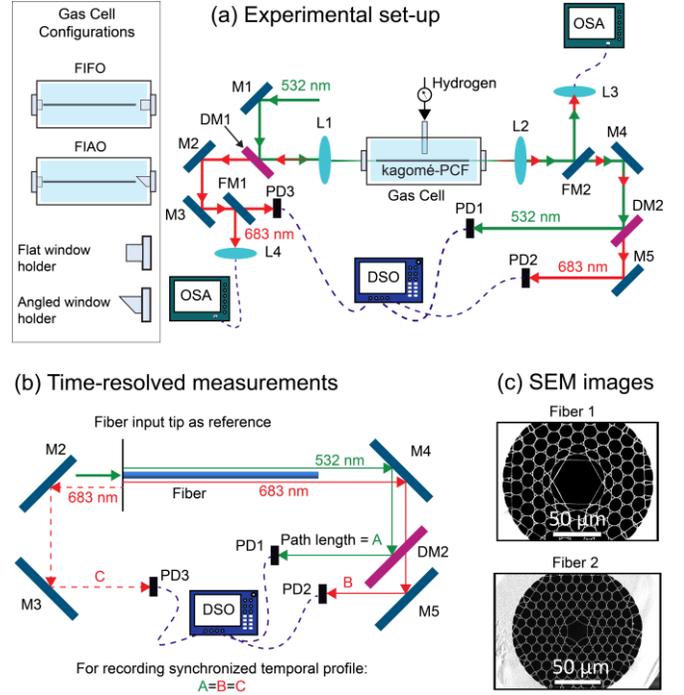

Fig. 1. Sketch of (a) the experimental set-up and (b) the arrangement of photodiodes for time-resolved measurements. (c) SEMs of the fibers used in the experiments with core diameters of ~47 μm (Fiber 1) and ~22 μm (Fiber 2). L: lens, M: mirror, DM: dichroic mirror, PD: photodiode, OSA: optical spectrum analyzer, DSO: digital storage oscilloscope.

## 4. Experimental results and discussion

To demonstrate and unambiguously distinguish self- and noise-seeded generation of the backward Stokes signal, two main gas-cell configurations were employed for both fibers (see Fig. 1(a)): (a) Flat input window and flat output window (FIFO); (b) Flat input window and angled output window (FIAO). Furthermore, for the experiments with the short fiber, an additional configuration was employed with angled input window and flat output window (AIFO).

When the uncoated flat output window is in place, a fraction of the noise-seeded forward Stokes pulse diverging from the output tip of the fiber undergoes Fresnel specular reflection and can be partially coupled back into the fiber (constituting a backward-propagating seed for BSRS, as we will see below). In contrast, in an angled output window configuration, most of the light reflected at the window is diverted away from the fiber axis, thereby strongly minimizing the coupling of the divergent Stokes signal back into the fiber and hence the potential seeding of BSRS. The experimental and simulation results for the two different HC-PCFs with the FIFO and FIAO configurations will be separately discussed below.

### A. BSRS seeded by back-reflected FSRS

In these experiments, pulses with ~28 μJ were launched into the 14-cm-long fiber with ~47 μm core diameter (Fiber 1 in Fig. 1(c)) and filled with 38 bar of $H_2$. The measured temporal profiles of the pump, forward and backward Stokes (averaged over 200 pulses) for the FIFO and FIAO configurations are shown in Fig. 2, along with numerical simulations obtained by solving a set of bi-directional Maxwell-Bloch equations (see

Supplement for details). Throughout this manuscript the time $t$ is defined such that $t = 0$ ns when the peak of the pump pulse enters the fiber input. We define time of flight $t_f$ as the time taken for the peak of the initial pump pulse, in the absence of SRS, to reach the output face of the fiber, i.e. the fiber length divided by the group velocity. For Fiber 1, $t_f \approx$ 0.45 ns and for Fiber 2, 3.8 ns.

It is evident that the backward Stokes signal is strongly suppressed in the FIAO configuration due to the absence of self-seeding, while it becomes about 12 times stronger in the FIFO configuration. Moreover, in the FIFO configuration, the backward Stokes signal appears earlier in time (see Fig. 2(a, b)), strongly suggesting that the forward Stokes signal is reflected back into the fiber at the output window, acting as a seed for BSRS and triggering a quicker onset of backward amplification.

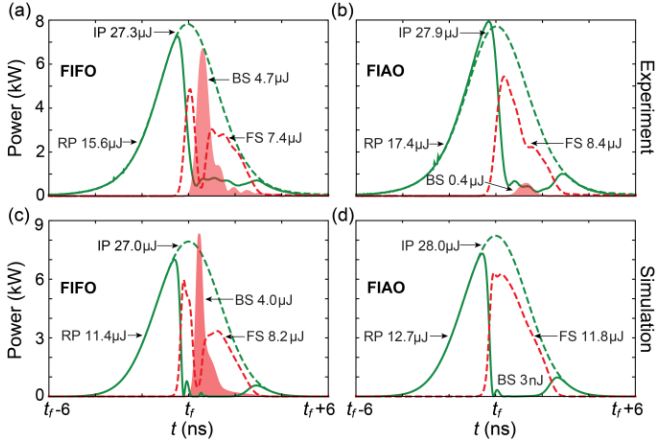

Fig. 2. (a,b) Experimental and (c,d) Simulated temporal profiles of the generated sidebands for the FIFO (left) and FIAO (right) configurations in Fiber 1 filled with 38 bar of $H_2$. The measured energies for the initial pump (IP), residual pump (RP), forward Stokes (FS) and backward Stokes (BS) are indicated in the panels.

Although the differences between the two configurations are clear, these measurements alone cannot be used to discriminate whether the weak backward Stokes signal observed in the FIAO configuration is fully noise-seeded, because back-reflection of light at the slanted output window, although strongly suppressed, cannot be completely excluded.

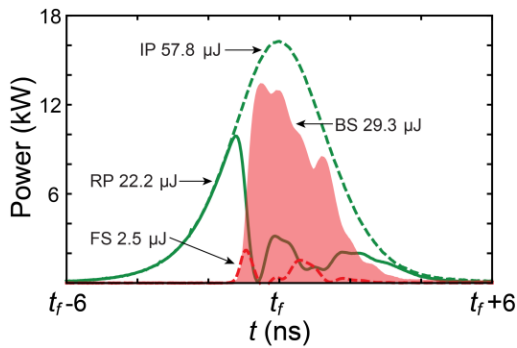

Fig. 3. Temporal profile of Raman sidebands when 58 µJ of pump pulse energy is launched into a 14-cm long Fiber 1, filled with 38 bar of $H_2$. We selected an AIFO configuration to enhance self-seeding. We see how at sufficiently high pump energies, the BSRS overtakes FSRS.

Interestingly, in the FIFO configuration the backward Stokes signal emerges in time between the first peak of the forward Stokes signal and its revival towards the trailing edge of the pump pulse. This double-peak structure (Fig. 2(a)) is also accurately reproduced in the simulations (Fig. 2(c)). Indeed, we have numerically corroborated that the threshold and revival of the FIFO forward Stokes signal (i.e., the second hump delayed by ~2 ns with respect to the peak of the initial pump) are influenced by the back-reflection of the backward Stokes signal at the input window of the gas cell (see Supplement). As a result, the backward Stokes signal is temporally-gated and compressed to 540 ps duration, with evident practical applications. Based on this understanding, we increased the launched pump energy to 58 µJ in the same fiber but in an AIFO geometry so as to inhibit the growth of the forward Stokes signal, and observed ~65% quantum conversion efficiency for backward Stokes generation (see Fig. 3) in a pure fundamental fiber mode.

**B. Noise-seeded BSRS**

To study the temporal dynamics of pure noise-seeded BSRS, we must first ensure that the pump pulse is significantly depleted before it reaches the output end of the fiber. This will avoid any back-seeding and therefore leave the FSRS process unaffected. To implement this we used a 114-cm-long fiber (longer than the FWHM length of the pump pulse) with smaller core diameter of ~22 µm (Fiber 2), again filled with 38 bar of hydrogen gas. The increased pump intensity in the smaller core and the longer fiber length dramatically reduce the SRS threshold, with the result that the pump is strongly depleted before reaching the fiber end. Under these circumstances the backward Stokes signal is most likely be generated solely from noise.

Using the FIAO configuration, we monitored the backward and forward Stokes signals, as well as the residual out-coupled pump light, while increasing the input pump energy (Fig. 4(a)). The data-points are the average over measurements from ten separate runs. The backward Stokes appears at a pump energy of ~3.1 µJ, thereafter growing at an average slope efficiency of ~40%. At the maximum launched energy of ~23.7 µJ, the backward Stokes energy was ~7.5 µJ, corresponding to ~32% energy conversion and ~41% quantum efficiency.

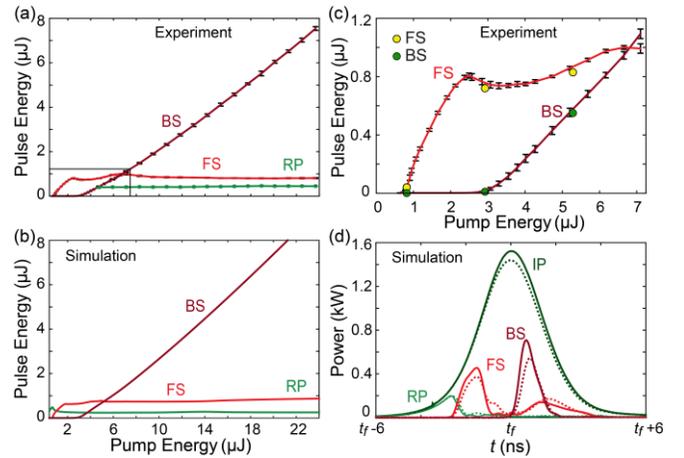

Fig. 4. (a) Experimental and (b) simulated output energies of residual pump, backward and forward Stokes signals with increasing pump energy. Fiber 2 was filled with 38 bar of $H_2$ in FIAO. The residual pump energy was not recorded for pump energies below 4 µJ. The region enclosed by a black box in (a) is shown in (c). The colored circles in (c) are data taken from experiments in the FIFO configuration. (d) Temporal profiles for a pump energy of 5.3 µJ. The solid and dashed lines represent the FIAO and FIFO configurations. IP: initial pump, RP: residual pump, FS: forward Stokes signal, BS: backward Stokes signal.

Although the gain for BSRS is lower than for FSRS (the threshold for FSRS is ~0.9 µJ pump energy), the backward Stokes signal clearly overtakes the forward Stokes signal at ~6.8 µJ of pump energy, beyond which the backward Stokes signal keeps growing while the forward Stokes signal and the residual pump light remain saturated at ~0.81 µJ and ~0.43 µJ respectively. Note that this saturation effect is also modified by conversion to higher-order Raman bands, both vibrational and rotational, and fiber losses. All these features are also confirmed by numerical simulations (see parameters in Supplement), which show excellent agreement with the experimental results (see Fig. 4(b)). In our simulations we employ a uniform low-amplitude field of 50 V/m as noise floor for all the lines (Supplement) to get a close agreement to the experiments. This accounts for processes such as spontaneous Raman scattering, vacuum fluctuations and initial thermal phonon population. Other effects such as Rayleigh scattering from the gas molecules and scattering from inner surfaces of the fiber-core wall may also add complexity to the determination of the noise floor when BSRS and FSRS processes influence each other. However simple theoretical considerations indicate that these effects are likely to become significant only at wavelengths much shorter than used in the current experiments, and can be safely neglected.

The numerical results are identical in both FIFO and FIAO configurations, pointing to the noise-seeded nature of the BSRS in this case. To further verify this experimentally, we measured the signals at pulse energies of 0.8, 2.9, 5.3 µJ in the FIFO configuration (marked as colored circles in Fig. 4(c)); they almost perfectly agree with those obtained in the FIAO configuration. This is in sharp contrast to the previous results for self-seeded BSRS, where the FIFO configuration was found to greatly enhance the generation of the backward Stokes signal. This is further confirmed by time-resolved measurements, as shown in Fig. 4(d). We believe the measurements represent the first unambiguous observation of efficient noise-seeded BSRS in hydrogen-filled HC-PCF.

The spectra of the forward and backward-propagating lines for ~24 µJ pump energy in Fiber 2 are shown in Fig. 5. The forward-propagating signal shows several Raman lines (see Fig. 5(a)), generated from both vibrational and rotational SRS. These may be explained by the long interaction length for co-propagating pump and Stokes/anti-Stokes lines, along with a shallow dispersion landscape that permits interactions with different forward coherence waves (which, among other effects, allows generation of up-shifted anti-Stokes signals via molecular modulation [21]). In contrast, the counter-propagating Stokes light is concentrated solely at 683 nm, corresponding to the first vibrational Stokes line (see Fig. 5(b)).

As already mentioned, this behavior may be understood by reference to the dispersion diagram in Fig. 6, which plots the frequency-dependent propagation constants $\beta^q$ of forward ($q = f$) and backward ($q = b$) propagating fundamental modes of Fiber 2 for 38 bar of $H_2$, based on the modified Marcatili-Schmelzer model (see Supplement). The solid arrows represent the 4-vectors of the coherence waves ($C_w$'s) generated by interference between pump and both forward and backward Stokes and anti-Stokes signals:

$$C_{wS}^q = (\beta - \beta_S^q, \Omega_R) \quad C_{wAS}^q = (\beta_{AS}^q - \beta, \Omega_R) \quad (2)$$

For efficient anti-Stokes generation, the momentum component of the coherence wave must be closely similar for Stokes and anti-Stokes processes. The dephasing rate can be written $\vartheta^q = |2b - b_{AS}^q - b_S^q|$, which works out at 3.1 rad/cm for FSRS and $4.8 \times 10^5$ rad/cm for BSRS, corresponding respectively to dephasing lengths $\pi/\vartheta^q$ of 1 cm and 65 nm. This strongly favors generation of a forward frequency comb, while very strongly suppressing generation of a backward comb. This explains the simplicity of the observed backward spectra and the strength of the backward Stokes signal.

Furthermore, the backward Stokes signal is always emitted in a clean fundamental core mode (see insets in Fig. 5(b) for a far-field profile). Very strong dephasing also means that BSRS is not impaired by coherent Raman gain suppression [15, 22] (see Supplement).

## 5. Competing forward/backward gain

In Section 4B we saw that the backward Stokes signal is always stronger than the forward at high $H_2$ pressure and high pump power, despite the higher forward Raman gain. This may be explained as follows. After the pump pulse enters the fiber, the forward Stokes signal emerges earlier in time and also with lower threshold than the backward due to its higher Raman gain (see Fig. 4). Simultaneously, the pump also gives rise to a small backward Stokes signal from noise-driven spontaneous scattering. This weak backward signal gets amplified as the pump pulse travels through the fiber, yielding to enhancement of the backward coherence wave. Meanwhile, the FSRS process continues uninterrupted until the backward Stokes build-up is strong enough to cause pump depletion at the leading edge and central part of the pulse. Nevertheless, the backward coherence wave, left behind once the backward Stokes signal has left the fiber via the input end, keeps aiding the scattering of further Stokes photons from the trailing edge of the pump pulse. Finally, at some point the backward signal decays, due to local depletion of the pump.

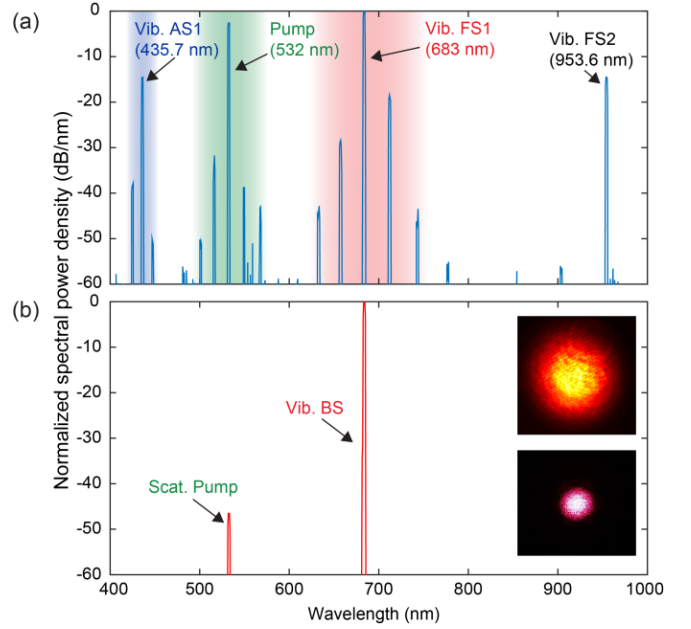

Fig. 5. Spectrum of (a) the forward and (b) the backward propagating pulses normalized to the first vibrational Stokes signal. A ~24 µJ pump pulse was launched into Fiber 2 in the FIAO configuration. The forward spectrum (a) consists of a hybrid ro-vibrational Raman comb. The shaded regions enclose the rotational lines for the respective pump or vibrational line. In sharp contrast, the backward spectrum (b) is very simple, containing only the vibrational backward Stokes BS and a small fraction of scattered pump. The upper inset in panel (b) is the far-field image of the BS signal imaged with a CCD camera and the lower inset a photograph of the BS signal cast onto a screen.

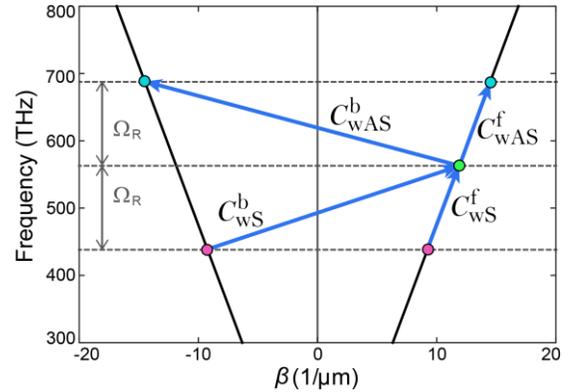

Fig. 6. Dispersion diagram for the forward and backward $LP_{01}$-like modes of the hollow-core PCF. The solid blue arrows represent the coherence waves involved in the various different SRS transitions. Backward anti-Stokes generation is very strongly dephased, as expected. FSRS is also dephased, but by a much smaller degree (too small to be seen on the plot). See the text for more details.

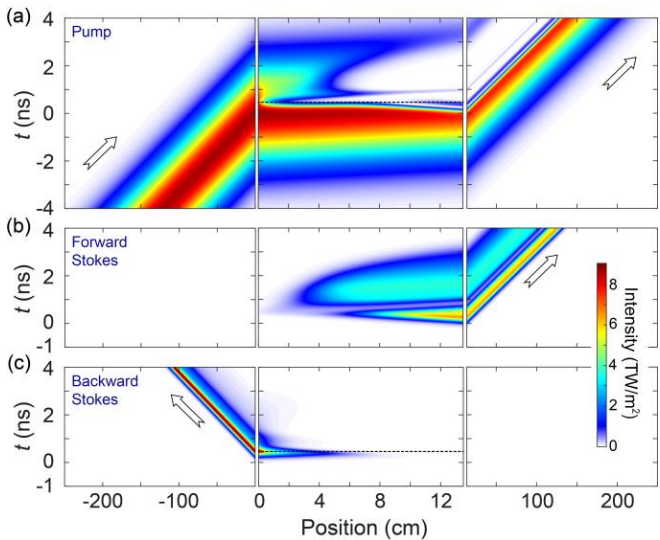

Fig. 7. Simulated spatio-temporal evolution of (a) the pump, (b) forward Stokes and (c) backward Stokes signals. The simulation parameters correspond to those used in Fig. 2(c). The dashed horizontal lines mark the time when the backward Stokes signal attains its peak intensity. Note that the spatial scale inside the fiber is magnified for clarity.

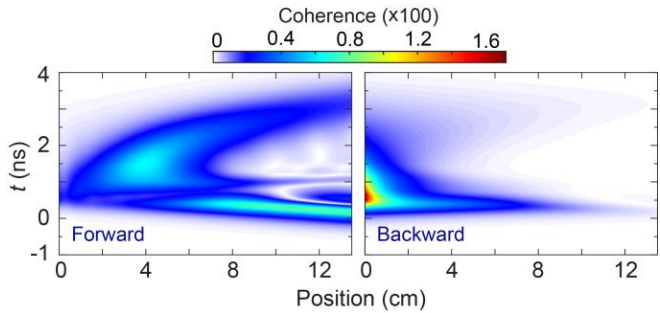

Fig. 8. Simulated spatio-temporal evolution of the forward and backward coherence waves. The parameters correspond to those used in Fig. 2(c) and Fig. 7. In spite of the lower overall Raman gain, the peak strength of the backward coherence is more than double the forward.

The remarkable agreement between experimental measurements and numerical simulations makes it possible to obtain additional information about the system by examining the competing dynamics of the forward and backward coherence. Figure 7 shows the simulated spatio-temporal evolution of the pump, forward and backward Stokes signals for the parameters used in the FIFO configuration in Fig. 2(c). The causal sequence of scattering events may be clearly traced. Conversion to the backward Stokes signal takes place within a very narrow space-time window, close to the entrance face of the fiber where the backward coherence wave is strongest (see Fig. 8). In contrast, the forward coherence (Fig. 8) peaks in the second half of the fiber, extending from ~8 to ~14 cm (the secondary forward coherence peak, centered at ~1.5 ns, is caused by reflection of the backward Stokes signal at the input of the fiber). The short backward interaction length means that the backward Stokes signal is unable to generate its own Stokes lines (see Fig. 5(b)), with the result that the Stokes energy is concentrated within a narrow spectral band, and also has the consequence that fiber losses play only a minor role, unlike for forward SRS.

## 6. Conclusions

In conclusion, backward Stokes light can be efficiently generated in a short length of gas-filled HC-PCF in a simple arrangement involving only a monolithic gas-cell. The system allowed observation for the first time of unambiguously noise-seeded BSRS in hydrogen-filled HC-PCF, with a quantum efficiency of 41%. When the backward SRS is self-seeded by back-reflected forward Stokes light, the quantum efficiencies can be as high as 65%, even though the forward Raman gain of the gas is higher than the backward. The efficiency of the effect will be even higher for pumping in the ultraviolet, when the Raman gain is much stronger [22]. The backward Stokes light is spectrally very narrow and has a high quality $LP_{01}$-like mode profile, while the backward Raman coherence is concentrated close to the input face of the fiber, preventing generation of higher-order Stokes side-bands. Together with recent developments in hollow-core fiber technology [23], the results pave the way to a new generation of fiber-based Raman lasers and amplifiers, ideal for operation in otherwise difficult-to-access spectral regions such as the ultraviolet and the mid-infrared. In preliminary non-optimized experiments, not reported here, we have demonstrated generation of deep-ultraviolet backward Stokes light at 299 nm from 266 nm pump with energy conversion efficiencies exceeding 10%. The HC-PCF system also provides a novel platform for ultrafast pulse compression and amplification in counter-propagating geometries [24].

See Supplement for supporting content.

# Supplementary material

## 1. Bi-directional Maxwell-Bloch equations

In our system involving pump pulses of few nanoseconds duration propagating in gas-filled hollow-core fibers, we simulate the coupled spatio-temporal dynamics of the forward (FSRS) and backward (BSRS) Raman scattering processes by means of a bi-directional version of the Maxwell-Bloch equations (BMBE) [1-3]. When only the fundamental core mode is considered, we solve the following set of coupled equations:

$$\left(\frac{\partial}{\partial z}+\frac{1}{c}\frac{\partial}{\partial t}+\frac{1}{2}\alpha_l\right)E_l^f = -i\kappa_{2,l}^f \frac{\omega_l}{\omega_{l-1}} Q_f E_{l-1}^f q_{l-1}^f q_l^{f*}$$
$$-i\kappa_{2,l+1}^f Q_f^* E_{l+1}^f q_{l+1}^f q_l^{f*}, l \neq 0 \quad \textbf{(S1)}$$

$$\left(\frac{\partial}{\partial z}+\frac{1}{c}\frac{\partial}{\partial t}+\frac{1}{2}\alpha_0\right)E_0^f = -i\kappa_{2,0}^f \frac{\omega_0}{\omega_{-1}} Q_f E_{-1}^f q_{-1}^f q_0^{f*}$$
$$-i\kappa_{2,1}^f Q_f^* E_1^f q_1^f q_0^{f*}$$
$$-i\kappa_{2,0}^f \frac{\omega_0}{\omega_{-1}} Q_b E_{-1}^b \quad \textbf{(S2)}$$

$$\left(-\frac{\partial}{\partial z}+\frac{1}{c}\frac{\partial}{\partial t}+\frac{1}{2}\alpha_{-1}\right)E_{-1}^b = -i\kappa_{2,0}^b Q_b^* E_0^f \quad \textbf{(S3)}$$

$$\frac{\partial}{\partial t}Q_f = -Q_f/T_2 - \frac{i}{4}\sum_l \kappa_{1,l}^f E_l^f E_{l-1}^{f*} q_l^f q_{l-1}^{f*} \quad \textbf{(S4)}$$

$$\frac{\partial}{\partial t}Q_b = -Q_b/T_2 - \frac{i}{4}\sum_l \kappa_{1,0}^b E_0^f E_{-1}^{b*} \quad \textbf{(S5)}$$

where $l$ is an integer representing the sidebands of the pump ($l = 0$) with frequency $\omega_P$ such that $\omega_l = \omega_P + 2\pi l \Omega_R$ where $\Omega_R$ is the Raman frequency shift. The complex electric fields for forward propagating lines are defined as $e_l^f(z,t) = E_l^f(z,t)q_l^f$ where $E_l^f(z,t)$ denotes the slowly-varying complex field envelope and $q_l^f = \exp(-i\beta_l^f z)$, where $\beta_l^f$ is the propagation constant of the individual Raman lines. $z$ is the propagation distance and $t$ is the absolute time. For the backward-propagating lines, the superscript $f$ is simply replaced by $b$. The losses for individual lines are denoted by $\alpha_l$. The coherence amplitudes for FSRS and BSRS are denoted by $Q_f$ and $Q_b$. The frequency-dependent coupling coefficients for FSRS are given by:

$$\kappa_{1,l}^f = \sqrt{\frac{2\gamma_f c^2 \varepsilon_0^2}{NT_2 \hbar \omega_{l-1}}} \quad \textbf{(S6)}$$

and

$$\kappa_{2,l}^f = \frac{N\hbar \omega_{l-1} \kappa_{1,l}^f}{2\varepsilon_0 c} \quad \textbf{(S7)}$$

where $N$ is the molecular number density, $\hbar$ the reduced Planck's constant, $c$ the speed of light in vacuum, $\varepsilon_0$ the vacuum permittivity and $T_2$ is the dephasing time of the Raman polarization, which is linked to the linewidth $\Delta \nu$ of the Raman transition by $\Delta \nu = 1/\pi T_2$. The coupling coefficients for the BSRS process can be obtained from those of FSRS taking into account the different material gain for the two processes. In particular, the gain for the FSRS $\gamma_f$ can be obtained from experimentally measured values [4], while the gain for the BSRS process is estimated by $\gamma_b = R\gamma_f$ where $R$ is the backward to forward gain ratio introduced in the main text. Here, we neglect the group velocity walk-off for the Raman lines and take the group velocity of individual lines to be $c$ in Eqs. S1-S3. These equations are valid in the limit where the majority of the molecules remain in the ground state.

The wavelength-dependent propagation constants $\beta(\lambda)$ of the modes of a gas-filled kagomé-PCF can be analytically approximated to good accuracy by the modified Marcatili-Schmelzer model [5]:

$$\beta(\lambda) = \sqrt{k_0^2 n_{gas}^2(p,\lambda) - u_0^2/a^2(\lambda)} \quad \textbf{(S8)}$$

where $k_0 = 2\pi/\lambda$ is the vacuum wavevector, $n_{gas}$ is the refractive index of the filling gas with pressure $p$, $u_0 = 2.4048$ and $a(\lambda) = a_{AP}/(1 + s\lambda^2/(a_{AP}d))$ where $a_{AP}$ is the area-preserving core radius, $s = 0.08$ is an empirical parameter, and $d$ is the core-wall thickness.

## 2. Arrangement of the photodiodes for time-resolved measurements

To gain more insight into the temporal dynamics of the BSRS process discussed in the paper, we performed accurate time-resolved measurements of the different pulse profiles using fast photodiodes. The arrangement of the different detectors to correctly and unambiguously investigate forward and backward dynamics simultaneously is, however, far from trivial. The basic idea of the arrangement of photodiodes and execution of the numerical modeling is based on the fact that the BSRS and FSRS processes start competing as soon as the pump pulse enters the input tip of the fiber (i.e., the interaction region). Subsequently, the backward Stokes moves out and the forward Stokes propagates further into the fiber—both from the input tip of the fiber. Hence, the input tip of the fiber is chosen to be equidistant from the photodiodes experimentally, and serves as a reference and center of the array in the numerical modeling, as it will be shown below.

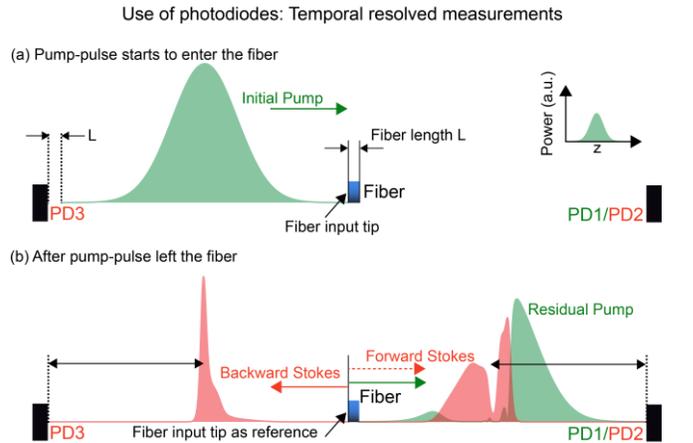

Fig. S1. Sketch of the arrangement of the photodiodes when the pump pulse (a) starts to enter the fiber and (b) has completely left the fiber. The green-solid, red-dashed, and red-solid arrows indicate the direction of propagation of pump, forward Stokes and backward Stokes, respectively. The double-headed black arrows in (b) are equal in length.

Figure S1 gives a schematic view of the arrangement of the photodiodes. The pulses shown in this figure are taken from the simulation results shown in Fig. 2(c) of the main text. Hence, Fig. S1(b) illustrates the scenario when the Raman interaction is over and the trailing edge of the pump-pulse has just left the fiber output. The length of the fiber in Fig. S1 is scaled according to the length of the pulses. The leading edge of the pump pulse passes through the gas-filled fiber without undergoing any changes (its amplitude is very low) and reaches the photodiode PD1. Simultaneously, the leading edge of the forward Stokes lies spatially in between the fiber output and the PD2—yet to reach PD2. Similarly, the leading edge of the backward Stokes signal is yet to reach PD3. The double-headed black arrows at PD1/2 and PD3 in Fig. S1(b), indicate equidistant points from these detectors. Upon comparison between Fig. S1 and Fig. 2(c) in the main text, one can see that the backward Stokes pulse indeed occupies the gap in between the two lobes of the forward Stokes.

## 3. Numerical modeling

In our simulations, four forward Raman lines were considered: first anti-Stokes, Pump, first Stokes and second Stokes, as well as one backward Raman line (first Stokes). For simplicity and based on the experimental results, only the fundamental core mode was considered. In addition to the modeling reported in [3], here we have included the onset of backward coherence waves. In particular, both forward $Q_f$ and backward $Q_b$ Raman polarization components are coupled through the pump (see Eqs. S1-S5). The numerical modeling mainly helps us to understand the temporal dynamics of the backward Stokes generation along with its competition with the forward lines, with the value for the backward/forward Raman gain ratio $R = 0.7$. As in the experiments, the hydrogen gas pressure has been fixed to 38 bar in the numerical modeling. The gain for the forward SRS has been incorporated from [4]. To get close agreement with the experiment, a uniform low-amplitude field of 50 V/m (noise floor) was chosen for all the lines and the pump pulse energies were varied within 10% of the experimentally recorded value.

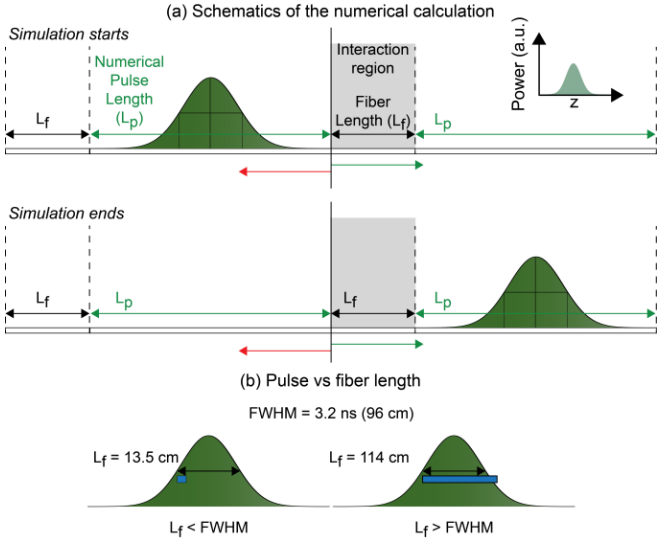

Fig. S2. (a) Starting and finishing conditions for the simulation. (b) Graphical comparison of the pump pulse length to the lengths of the two fibers employed in the experiments.

**Defining the numerical arrays:** a time step $\Delta t = 1$ ps was chosen as a compromise between the computational time and the convergence of the solutions obtained by numerically solving the BMBE. In this method, the longitudinal spatial step along the fiber axis is then univocally fixed as $\Delta z = \Delta t c$. The total spatial length of the numerical arrays containing the information of the different sidebands is $2L_p + 2L_f$, where $L_p$ is the physical length of the pump pulse (defined as the full-width of the pulse where the intensity drops to ~$10^{-4}$ of its maximum) and $L_f$ is the fiber length.

The pump pulse is defined as the initial condition in a 12 ns-long array and placed at the appropriate position with respect to the fiber input end (as shown in Fig. S2, the leading edge of the pump-pulse starts just before the input tip of the fiber ($L_f + \Delta z$ to $L_f + L_p$). The rest of the pump pulse array is filled up with the low-amplitude noise floor.

**Numerical evolution:** once the initial conditions are set, we consider that the pump pulse propagates from left to right as shown in Fig. S2, which implies that the forward-propagating lines will also travel from left to right, and the backward Stokes will do the opposite. The evolution ends when the trailing edge of the pump pulse, lying initially at $L_f + \Delta z$, leaves the fiber and reaches the point $L_f + L_p + L_f + \Delta z$. As a result, at the end of the simulation, all the temporal information for the forward lines is contained between $2L_f + L_p$ and $2L_f + 2L_p$ while the corresponding information of the backward Stokes lies in between 0 and $L_f + L_p$. To be compared with the experiments, the numerical temporal profiles are then temporally mapped onto one another to give the final results. In this process, the input tip of the fiber acts as the reference point.

**Self-Seeding:** To incorporate the self-seeding processes, the model must rigorously account for the reflections from the gas-cell windows at normal incidence. In this regard, back-reflections are implemented by multiplying at each computational step the estimated back-reflection coefficient with the forward/backward Stokes field at the window, and subsequently adding up the result to the uniform noise amplitude of the corresponding sideband at the position of the input/output window.

The details regarding the calculation of the back-reflection and its subsequent coupling into the fiber are given below.

## 4. Modeling back-reflections at the gas-cell windows

The back-reflection coefficients for the gas-cell windows were calculated using the following procedure:
1. Based on Fresnel theory, a normal-incidence reflection coefficient $R_c \approx 0.05$ is obtained for the MgF$_2$ windows at 683 nm.
2. We analytically approximate the LP$_{01}$-like transverse electric field distribution of the Stokes as field:

$$E_{-1}(x, y, 0) = A J_0(u_0 r / a_{AP}) \quad \textbf{(S9)}$$

where $A$ is the electric field amplitude, $r = \sqrt{x^2 + y^2}$ is the radial coordinate and $J_0$ is the zero-order Bessel function of the first kind.
3. The input and output windows of the gas cell are placed at a known distance $d$ away from the input/output tip of the fiber. Thus, the diffracted beam $E_{-1}(x, y, z = 2d)$ is calculated by using a standard Fourier-based beam propagation method. As the beam propagates in free space from the fiber tip to the window and back, it covers an effective distance $2d$.
4. The power coupling efficiency $K$ between the LP$_{01}$-like fiber mode and its diffracted signal over a distance $2d$ is calculated using the integral [6]:

$$K = \frac{\left(\iint \left| E_{-1}(x,y,0) \times E^*_{-1}(x,y,2d) \right| dxdy \right)^2}{\iint \left| E_{-1}(x,y,0) \right|^2 dxdy \times \iint \left| E_{-1}(x,y,2d) \right|^2 dxdy} \quad \textbf{(S10)}$$

5. Finally, the back-reflection coefficient is calculated as $\sqrt{K} \times \sqrt{R_c}$.

## 5. Additional parameters of the experiments

In this section, we include additional parameters used for modeling of the two sets of experiments discussed in the main text.

Set 1: 13.5 cm long fiber with 47 µm core diameter
For this short fiber length, no propagation losses were included in the simulation. The approximate position (and calculated back-reflection coefficient) of the input and the output window for the FIFO and FIAO configurations are 2 mm (0.1) and 5 mm (0.043) respectively. For the FIAO configuration, the back-reflection from the output window is switched off.

Set 2: 114 cm long fiber with 22 µm core diameter
In the experiments with the longer fiber, we observed several rotational lines appearing in the spectra. However, given that they were in general more than 20 dB weaker than the first vibrational Stokes, we restricted our modeling only to vibrational Raman lines for simplicity.

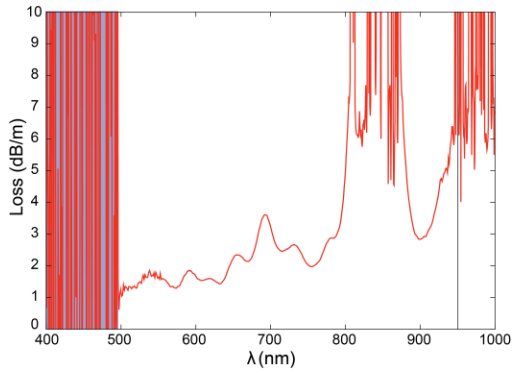

Fig. S3. Loss measurement for the 22 µm core fiber. There are no valid data below 500 nm, the limit of the supercontinuum source employed in the measurements. The vertical solid line at 953.6 nm indicates that the second Stokes lies in a high-loss band.

In this case we also considered propagation losses for the different sidebands. Based on the measured loss of the fiber (see Fig. S3), the losses of the pump (532 nm) and first Stokes (683 nm) are taken as 1 and 3 dB/m respectively. The loss for the first anti-Stokes (435.7 nm) is a free parameter due to the lack of available data below 500 nm (we have chosen 1 dB/m based on experimental observation). The second Stokes (953.6 nm) lies in a high-loss window of the fiber, which also makes the prediction of its propagation constant based on the resonance-free capillary model (Eq. S8) invalid. Hence, for a better agreement to experiments, the propagation constant for this line was increased by 120 rad/m compared to the value calculated from equation S8 [5] and the loss value was taken as 45 dB/m.

In this set of simulations, the approximate position (and calculated back-reflection coefficient) of the input and the output window for the FIFO and FIAO are 7 mm (0.007) and 5 mm (0.01) respectively. Again, for the FIAO configuration the back-reflection from the output window is switched off.

## 6. Forward Stokes influenced by the reflection of backward Stokes from the input window of the gas-cell

In the main text, Fig. 2(a, c) show the experimental and simulation results for the FIFO configuration. There we point out that the back-reflection of the backward Stokes at the input window of the gas-cell influences both the threshold and the double-hump temporal structure of the forward Stokes. Figure S4 shows the numerical result of a situation similar to that depicted in Fig. 2(c) in the main text, but with the input back-reflection coefficient being numerically switched off (i.e., mimicking an AIFO configuration). Under these conditions the forward Stokes signal is dramatically suppressed and exhibits only a single lobe.

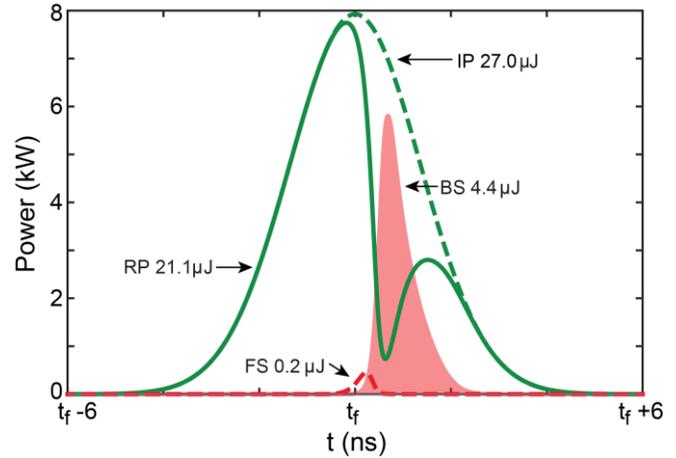

Fig. S4. Simulation including an AIFO configuration for the parameters of Fig. 2(c) in the main text.

## 7. Near-field mode profiles in the gain-suppression regime

In order to experimentally determine the pressure range over which coherent Raman gain suppression effects are noticeable in the ~22 µm-core fiber, we monitored the near-field profile of the forward first Stokes signal while changing the $H_2$ pressure [3]. By launching 3 µJ of pump energy and observing the transition of the modal content of the Stokes signal (see Fig. S5) from an $LP_{01}$-like to $LP_{11}$-like mode and back, we verified that the Raman gain suppression region spans from 18 to 21 bar. As it has been already reported, the gain suppression region will broaden with increasing input energy [7]. In any case, as all the experiments presented here were carried out with 38 bar of $H_2$, the potential influence of gain suppression effects in our system is minimal.

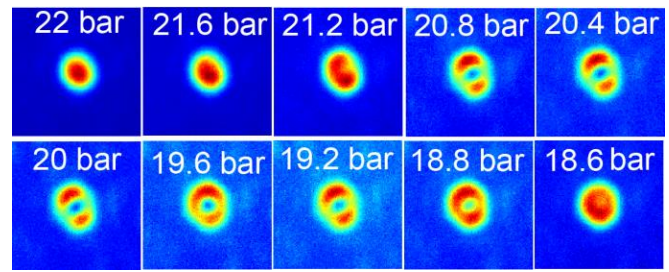

Fig. S5. Near-field profiles of the forward Stokes line at 683 nm recorded at various hydrogen pressures around the coherent gain suppression region. The fiber core diameter was 22 µm and the input energy 3 µJ.